\documentstyle[epsf]{l-aa}
\newcommand{\etal}{et~al.\ }

\title{BeppoSAX follow-up search for the X-ray afterglow of GRB970111}
\author{M. Feroci\inst{1} 
\and L.A. Antonelli\inst{2}
\and M. Guainazzi\inst{2}
\and J.M. Muller\inst{2,3}
\and E. Costa\inst{1}
\and L. Piro\inst{1}
\and J. Heise\inst{3}
\and J.J.M. in 't Zand\inst{3}
\and F. Frontera\inst{4,5}
\and D. Dal Fiume\inst{4}
\and L. Nicastro\inst{4}
\and M. Orlandini\inst{4}
\and E. Palazzi\inst{4}
\and G. Zavattini\inst{5}
\and P. Giommi\inst{2}
\and A.N. Parmar\inst{6}
\and A. Owens\inst{6}
\and A.J. Castro-Tirado\inst{7}
\and M.C. Maccarone\inst{8}
\and R.C. Butler\inst{9} 
}

\begin{document}
\offprints{feroci@saturn.ias.rm.cnr.it}
\date{Received 13 January 1998; accepted 3 March 1998}
\thesaurus{011 (13.07.1; 13.25.1)}

\institute{{Istituto di Astrofisica Spaziale, CNR, 
Via Fosso del Cavaliere, I-00133 Roma, Italy}
\and
{BeppoSAX Scientific Data Center, Via Corcolle 19, I-00131 Roma,
Italy}
\and
{Space Research Organization in the Netherlands, Sorbonnelaan 2,
3584 CA Utrecht, The Netherlands}
\and
{Istituto Tecnologie e Studio Radiazioni Extraterrestri, CNR, 
Via Gobetti 101, I-40129 Bologna, Italy}
\and
{Dipartimento di Fisica, Universit\`a di Ferrara, Via Paradiso 11, 
I-44100 Ferrara, Italy}
\and
{Astrophysics Division, Space Science Department of ESA, ESTEC,
P.O. Box 299, 2200 AG Noordwijk, The Netherlands}
\and
{Laboratorio de Astrofisica Espacial y Fisica Fundamental, INTA,
Madrid, Spain}
%\and
%{Istituto Fisica Cosmica e Tecnologie Relative, CNR, Via Bassini 15,
%I-20133 Milano, Italy}
\and
{Istituto Fisica Cosmica e Applicazioni Informatica, CNR, Via U. La Malfa
153, I-90146 Palermo, Italy}
\and
{Agenzia Spaziale Italiana, Viale Regina Margherita 202, I-00162 Roma,
Italy}
}

\maketitle

\begin{abstract}
The BeppoSAX satellite has recently opened a new way towards 
the solution of the long standing gamma-ray bursts' (GRBs) enigma, providing 
accurate coordinates few hours after the event 
thus allowing for multiwavelength follow-up observational
campaigns. 
%In this paper we present the first rapid follow up 
%observation performed by BeppoSAX itself on the error box of a GRB. 
%A Target of Opportunity observation with the 
%sensitive X-ray telescopes aboard BeppoSAX  
%started only 16 
%hours after that GRB970111 triggered its Gamma Ray Burst Monitor
%and that the Wide Field Cameras accurately 
%located the event. This observation has lead
%to the detection of an unknown X-ray source, 1SAX J1528.1+1937,
%positionally compatible with the GRB error box given by the WFC, but 
%inconsistent with the combination of the above error box with
%the IPN annulus.
%
%Whether or not the new source 1SAX J1528.1+1937 is associated
%to the GRB970111, it exhibited a flux level much lower than expected
%in case the same phenomenon operates in each and every GRB.
%In fact, given that GRB970111 is the brightest BeppoSAX
%GRB, its case shows explicitly that there is
%no obvious relation between the GRB gamma-ray peak flux and 
%the presence/intensity of the afterglow.
%In the framework now set up by BeppoSAX 
%this detection is of key importance for a general comprehension 
%of the GRB phenomenon.
%
% Suggestion by A. Parmar:

The BeppoSAX Narrow Field Instruments observed the region of sky
containing GRB970111 16 hours after the burst. In contrast to other GRBs
observed by BeppoSAX no bright afterglow was unambiguously observed. 
A faint source (1SAXJ1528.1+1937) is detected in a position consistent
with the BeppoSAX Wide Field Camera position, but unconsistent with the
IPN annulus. Whether 1SAXJ1528.1+1937 is associated with GRB970111 or 
not, the X-ray intensity of the afterglow 
is significantly lower than expected, based on the 
properties of the other BeppoSAX GRB afterglows. Given that GRB970111
is one of the brightest GRBs observed, this implies that there is no
obvious relation between the GRB gamma-ray peak flux and the
intensity of the X-ray afterglow.


\end{abstract}

\section{Introduction}

The comprehension of the nature of the Gamma-Ray Bursts (GRBs)  
is a long-standing problem of a
world-wide scientific community since
the announcement of their discovery   (Klebesadel \etal 1973).
Many observational (Fishman \& Meegan 1995)
and theoretical (Lamb 1995; Paczynski 1995) efforts 
did not succeed in understanding the origin of GRBs. 
%in finding the breakthrough to the final solution.
The launch of the BeppoSAX satellite (Boella \etal 1997a) 
revolutionized the field, opening a new observational window 
soon after the GRB event.
Due to its Gamma Ray Burst Monitor (GRBM, 40--700~keV, 
Frontera \etal 1997a; Feroci \etal 1997a) and its Wide Field 
Cameras (WFCs, 2--26~keV, 
Jager \etal 1997) this satellite is capable of detecting GRBs 
in the gamma-ray band and accurately localizing them in 
X-rays through a coded mask proportional counter. 
%Since the beginning of 1997, at a gross rate 
%of one per month, BeppoSAX has distributed GRB positions 
%with 3 arcmin error radius just few hours after the main event, and
%has performed Target of Opportunity observations
%with its high sensitivity X-ray telescopes.

Five GRBs, amongst those simultaneously detected by the
GRBM and the WFCs, were promptly analyzed, allowing    
multiwavelength follow-up observational campaigns. The first 
result is the BeppoSAX discovery of the X-ray afterglow 
of GRB970228 (Costa \etal 1997, Costa \etal 1997a) 
and the discovery of a related 
optical transient by ground-based telescopes  
(van Paradijs \etal 1997).
Further results have been achieved with the detection of the
X-ray afterglows of GRB970402 (Feroci \etal 1997b, Piro \etal 1997a), 
GRB970508 (Costa \etal 1997c, Piro \etal 1997b) and GRB971214 
(Heise \etal 1997a, Antonelli \etal 1997). From GRB970508 an indication 
of an extragalactic origin has been derived through
the detection of an optical transient (Bond, 1997; 
Djorgovski \etal 1997)
and the measurement of its optical spectrum (Metzger \etal 1997),
providing a lower limit of 0.835 for the redshift of the possible GRB 
optical afterglow.

One of the most intriguing mysteries of GRB emitters
is possibly solved, but the overall picture is far from clear. 
In fact, out of the five events for which BeppoSAX performed rapid
follow up searches of a X-ray counterpart, one (GRB970111) has given a
result that is significantly different from the other four. 
The celestial location of GRB970111 was observed by BeppoSAX
just 16 hours after the GRB event, and no unambiguous evidence 
for an X-ray afterglow was found. 
A new faint source (1SAX J1528.1+1937)
was detected at a flux level that is much lower 
than that expected on the basis of the properties of the other 
GRBs later observed by BeppoSAX. 
Here we present this detection, discuss its association 
with GRB970111 and the diversity from 
the other four BeppoSAX GRBs.

\section{GRBM and WFC detection}

On 1997 11 January, 09:43:59.99 UT the GRBM onboard BeppoSAX
was triggered by an intense gamma-ray burst, showing a multipeak
structure and lasting $\sim$43 s (Costa \etal 1997b). 
The peak intensity was $(3.9\pm0.3)\times 10^{-6}$~erg~cm$^{-2}$~s$^{-1}$ 
in the energy range 40--700~keV.
This GRB was also detected by the WFC unit 2,
with a similar time profile structure 
but a longer duration (see Fig. 1).
The 2--10~keV peak flux was 
$(4.1\pm0.7)\times 10^{-8}$~erg~cm$^{-2}$~s$^{-1}$.
The fluence of the event in 40--700~keV was 
$(4.14\pm0.31)\times 10^{-5}$~erg~cm$^{-2}$ 
while in 2--10~keV it was $(1.6\pm0.1)\times 10^{-6}$~erg~cm$^{-2}$.  
In Fig. 1 the gamma-ray (GRBM) and X-ray (WFC) light curves 
of the event are shown.

\begin{figure}
%\picplace{7cm}
\epsfxsize=\hsize   \centerline{\epsffile{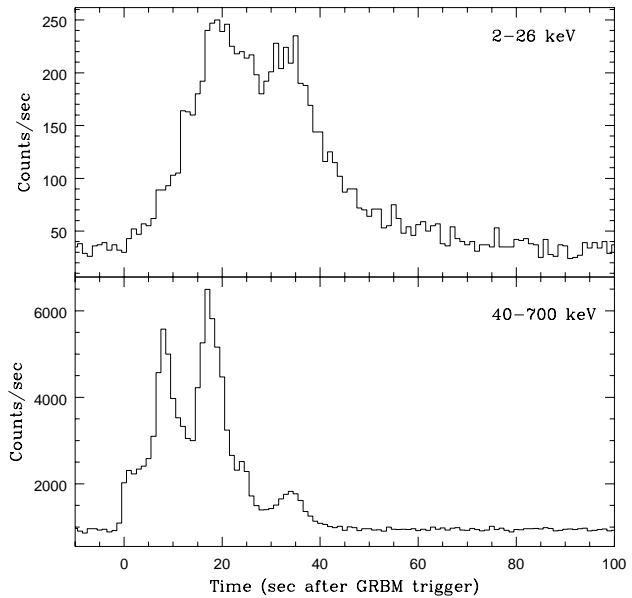}}
\vspace{-0.5cm}
\caption[]{BeppoSAX GRBM (40--700~keV) and WFC (2--26~keV) 
light curves of GRB970111}
\label{LC}
\end{figure}

Given that GRB970111 was one of the earliest X-ray transients 
detected in the    
WFC at the Quick Look Analysis, the position of the event was promptly
distributed with a 10$'$ error radius (Costa \etal 1997b), somewhat worse
than that obtainable from the instrinsic capabilities of the WFCs.
After about 20 days a revised error box of
the GRB970111 location with a 3$'$ error radius was produced, 
centred at a position 
4$'$.2 apart from the centroid of the previous one (in 't Zand \etal 1997). 
The new position was
R.A. = $15^{\rm h}28^{\rm m}15^{\rm s}$ and 
Decl.= +19$^{\circ}$36$'$.3 (equinox 2000.0).
Very recently the WFC hardware team again improved the instrument 
calibration, further reducing the error box area to an irregular
circle of 1$'$.8 radius (99\% confidence) (Heise \etal 1997), centred at
R.A. = $15^{\rm h}28^{\rm m}11^{\rm s}$ and Decl.=+19$^{\circ}$35$'$.9. 
This new region is contained within the previous one, but is centred
about 1$'$ apart. 

The Interplanetary Network used the delay in the GRB arrival times
between the interplanetary Ulysses mission and the  
ComptonGRO and BeppoSAX near-Earth 
satellites (Galama \etal 1997) to obtain a 
narrow strip of possible 
arrival directions in the sky. This allows  
the reduction of the GRB error box to a portion of the WFC error circle.

\section{BeppoSAX detection of 1SAX J1528.1+1937}

The earliest (10$'$) error box of GRB970111 was imaged with the narrow field
X-ray instruments (NFI) 
onboard BeppoSAX: the Low Energy Concentrator Spectrometer (LECS,  
0.1--10~keV, Parmar \etal 1997) and the three Medium Energy Concentrator
Spectrometers (MECS, 2--10~keV, Boella \etal 1997b). This Target of  
Opportunity observation was started 59,400 s after the GRBM  
trigger time, from 12 January 02:14 to 13 January 06:01 UT, for a 
total net exposure time of 52,139 s with the MECS and 11,594 s with the 
LECS (the latter being operated only during satellite night-time).

At the time when the NFI observation was performed 
the WFC improved error box was not available and therefore any
source included in the error box region was a potential counterpart
for the GRB970111. 
Two relatively bright X-ray sources were detected in 
the 10$'$ error box by the BeppoSAX/NFI
(Butler \etal 1997), resolved into three sources in the ROSAT All
Sky Survey (Voges \etal 1997). One of them, RXJ152845+1944.5, 
was also inside the early intersection
of the WFC error box and the IPN error strip (Hurley \etal 1997).
A peculiar radio source was detected with the VLA (Frail \etal 1997)
in a position coincident with this X-ray source.
The final WFC error box, however, excludes  
this source as possible counterpart to GRB970111. 

The latest WFC error box only includes an unknown X-ray source, 
1SAX J1528.1+1937.
The false colour image obtained from the MECS is shown
in Fig. 2. The image shows the
WFC error box, intersected by the IPN annulus, together with the 
MECS source error box.  

\begin{figure}
%\picplace{7cm}
\epsfxsize=\hsize   
\centerline{\epsffile{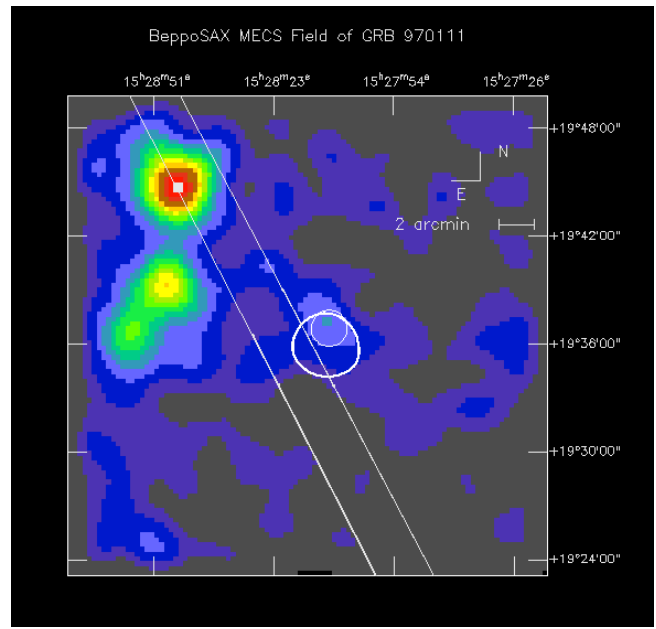}}
%\vspace{-3.5in}
\caption[]{False colour image of the MECS observation of the GRB970111. The
WFC error box is shown as an irregular circle containing the MECS error
circle of the X-ray source 1SAX J1528.1+1937. The latter lies outside the 
region of the WFC error box intersected by the IPN annulus. On the top 
left the source RXJ152845+1944.5 is clearly detected}
\label{fig:NFI}
\end{figure}

The data analysis of the MECS image, performed with the 
{\it Ximage} (Giommi \etal 1991), gives the position of the new source at
RA=$15^{\rm h}28^{\rm m}09^{\rm s}.2$ and Decl.=+19$^{\circ}$37$'$02", 
with a 60" error radius (90\% confidence level). 
The probability that this detection
is due to a background fluctuation is of the order of $10^{-6}$. 
The count rate of 1SAX J1528.1+1937 is
$(1.8\pm0.5)\times10^{-3}$~counts~s$^{-1}$ in the
MECS (2--10~keV) and $(7\pm3)\times10^{-4}$~counts~s$^{-1}$ in 
the LECS (0.1--2~keV).
Taking into account the vignetting correction for the off-axis
position, and assuming a Crab-like energy spectrum, 
the above count rates correspond to fluxes of
$(1.2\pm0.3)\times 10^{-13}$~erg~cm$^{-2}$~s$^{-1}$
in the 2--10~keV energy range and
$(8\pm4)\times 10^{-14}$~erg~cm$^{-2}$~s$^{-1}$ in the
0.1--2~keV energy range.

The MECS error box of the new source 
1SAX J1528.1+1937 is almost entirely contained within 
the WFC error box of GRB970111. 
Considering the 99\% confidence IPN annulus the source 
1SAX J1528.1+1937 is at a  position only marginally 
consistent with GRB970111. Therefore, if we use the reduced WFC-IPN
error box, the upper
limit to the 2--10~keV flux of
$1.6\times 10^{-13}$~erg~cm$^{-2}$~s$^{-1}$ (3 $\sigma$).

In the context of the possible association of 1SAX J1528.1+1937
with GRB970111 it is interesting to note that dividing the NFI
observation into three time intervals consisting of the first 10 ks,
the following 15 ks, and the last 26 ks of exposure time, gives the
count rates listed in Table 1. 
Even if the count rate is rather low, considering 
the combination of the temporal
and positional coincidences and the indication of
a decaying behavior, then the possibility of a random
occurrence of 1SAX J1528.1+1937 in the error box of GRB970111 is 
higher than the $\sim$3\%
derived from the
source statistics of the ASCA GIS (Cagnoni \etal 1997).

\vspace{-0.7cm}

\begin{table}
\label{tab}
\begin{center}
\caption{2--10~keV flux variation of 1SAX J1528.1+1937. The origin of the
elapsed time is the GRB970111 trigger time}
\begin{tabular}{|l|l|l|}
 \hline
Elapsed   & Count rate  & Flux(2--10~keV) \\  
Time (s)  & ($10^{-3}$~c~s$^{-1}$) & ($10^{-13}$~erg~cm$^{-2}$~s$^{-1}$) \\ \hline
59400-79700 & $(2.8\pm1.2)$  & $1.9\pm0.8$\\ \hline
79700-109940 & $(2.5\pm0.9)$  & $1.7\pm0.6$\\ \hline
109940-162530 & $(0.8\pm0.6)$  & $0.5\pm0.4$\\ \hline
\end{tabular}
\end{center}
\end{table}
\vspace{-1cm}

\section{Discussion and conclusions}

The BeppoSAX follow-up observation of the error box
of GRB970111 was the first prompt follow-up observation of a GRB
ever performed by an X-ray satellite.
Before BeppoSAX the time-scale of a possible X-ray emission 
from GRB remnants was completely unknown. 
This first basically non-detection, therefore, 
could only be interpreted as an upper limit to the 
time-scale of the decline of an X-ray afterglow or to its flux. 
Now, with the detection of the X-ray afterglows of GRB970228, 
GRB970402, GRB970508 and GRB971214, 
BeppoSAX has set up a new scenario in which GRB970111 seems misplaced. 
Also the detection of the X-ray afterglow of a GRB (GRB970828, 
Remillard \etal 1997; Murakami \etal 1997)
by the RossiXTE and the ASCA satellites  
supports the general framework for the GRBs' afterglow built by BeppoSAX. 

GRB970228, GRB970402 and GRB970828 showed a similar behavior, with a
fading X-ray counterpart continuously decaying from the GRB main
emission into the afterglow following an approximate $t^{-1.3}$ law.
In the case of GRB970228, the spectral analysis
(Frontera \etal 1997b) confirms the continuity between
the latest GRB emission and the X-ray counterpart
detected after few hours.
This temporal behaviour could be explained in the framework of 
the fireball model 
(Cavallo \& Rees 1978; Rees \& Meszaros 1992) as a 
highly radiative expansion of a relativistic  
shell. 
GRB970508 has shown a X-ray counterpart whose decay 
is more complicated than 
the above three. The above model could still account 
for this different behavior, but it needs to invoke a non-uniform 
surrounding medium, with a density scaling as $r^{-2}$  (Vietri 1997).

Whether 1SAX J1528.1+1937 is related to GRB970111  
or not, this gamma-ray burst had a much faster decay than
observed for any of the others. In order to make this clear, we compare
a hypothetic power-law decay of GRB970111 with the 
``typical'' power-law decay of GRB970228 reported in Costa \etal (1997a).
Therefore, in Fig. 3 we assume that the 
new X-ray source is associated with the GRB and impose a 
power-law decay of the afterglow starting from the WFC 
mean flux at a time centred on the GRB X-rays duration. 
The needed power-law index is -1.5. 

\begin{figure}
%\picplace{7cm}
\epsfxsize=\hsize   \centerline{\epsffile{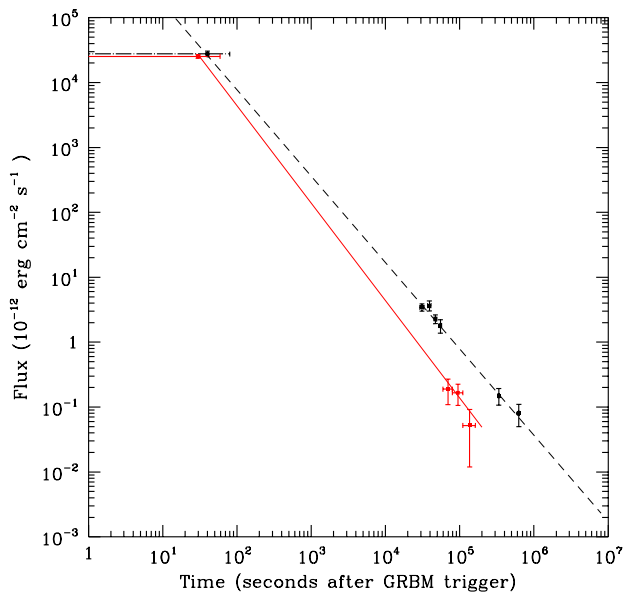}}
\vspace{-0.5cm}
\caption[]{X-ray (2--10~keV) 
decay law of the candidate counterpart of GRB970111,
compared to GRB970228. The dot-dashed and the solid horizontal 
lines are the mean X-ray flux for GRB970228 and GRB970111, respectively. 
The inclined dashed line is the decay law suggested by 
Costa et al. (1997a) for GRB970228. The inclined solid line shows
the power-law index, 1.5, needed for connecting the WFC GRB970111 mean flux
and the 1SAXJ 1528.1+1937 flux}
\label{fig:Decay}
\end{figure}

Alternatively, assuming that 1SAX J1528.1+1937 is not related 
to GRB970111 we can derive a  
lower limit to the power-law index by using the upper
limit of the MECS flux in the region of sky defined by the error
box, to obtain a value very similar to the 1.5 value given above.

Trying to extract GRB970111 from the group as an intrinsically 
different event, we note that its
gamma-ray fluence is about more than three times larger than 
the largest of the other three. On the other hand, even if this GRB
is of the ``No High Eenergy'' type (that is, it shows only weak emission
above 300 keV, Pendleton et al. 1997), the ratio
between the X-ray (2--10~keV) and gamma-ray (40--700~keV) 
fluences is about 4\%, to be 
compared to 20\% (2--10~keV) for GRB970228 (Frontera \etal 1997b),
5\% (2--10~keV) for GRB970402 (Nicastro \etal 1997) 
and 40\% (2--26~keV) for GRB970508 (Piro \etal 1997b).
GRB970111 appears therefore as the one (together with the April event) 
with the less efficient low X-rays
mechanism for energy release.
Furthermore, no optical source was found in the WFC error box
changing its intensity more than 0.5 
magnitudes at a level  of B=23 and R=22.6 from about 19 hours 
to about one month later 
(Castro-Tirado \etal 1997; Gorosabel \etal 1998). A radio search 
at 1.4 GHz (Frail \etal 1997) and at millimetric wavelength
(Smith \etal 1997) did not find a counterpart to 1SAXJ1528.1+1937.
These results support the idea that the optical, radio and
millimetric channels are unefficient as well.
Since GRB970111 was one of the brightest events ever
detected in gamma-rays, one may conclude that its gamma-ray channel was 
efficient enough to dissipate most of the energy generated in the
burst.

Alternative interpretations of the lack of X\-/optical\-/radio 
afterglow of the GRB970111 may be either
a very rapidly evolving afterglow, with a decay law faster than 
observed in the other BeppoSAX GRB afterglows, or the 
absence of an afterglow source. The former hypothesis would be in agreement
with the model by Tavani (1997) of a decay behavior represented
by a power law with index $-21/8$ due to the observation in a
fixed energy band (2--10~keV) of a synchrotron emission spectrum with a
rapidly evolving critical frequency. 
Alternatively, the latter situation could be due, as
an example, to the scenario in which the event that caused the
GRB occurred in a region in which the interstellar medium density is 
low enough (perhaps the external regions of a host
galaxy) to justify the absence of an external shock, possibly
responsible for the afterglow emission in the other cases 
(Katz \& Piran 1997).

%Any interpretation, however, will also have to take into account the peculiar
%energy dependence of the light curve of GRB970111. As it was shown
%in figure 1 the GRB emission at the X-ray energies is significant
%differently shaped from the gamma-ray light curve.
%Also the gamma-ray light curve provided by BATSE (Galama \etal 1997) in 4
%energy channels supports the strong energy evolution
%of the GRB from the gamma-rays to the X-rays. In particular 
%GRB970111 shows only a weak emission above 300 keV, that classifies
%it a "No High Energy" type (Pendleton \etal 1997).

\begin{acknowledgements}
This research is supported by the Italian Space Agency (ASI) and 
Consiglio Nazionale delle Ricerche (CNR). BeppoSAX is a major program 
of ASI with participation of the Netherlands Agency for Aerospace 
Programs (NIVR). All authors warmly thank the extraordinary teams 
of the BeppoSAX Scientific Operation Center and Operation Control 
Center for their enthusiastic support to the GRB program.
\end{acknowledgements}

\end{document}